\pdfoutput=1

\documentclass[sigconf,screen=true,natbib]{acmart} 
\copyrightyear{2019}
\acmYear{2019}
\setcopyright{iw3c2w3}
\acmConference[WWW '19]{Proceedings of the 2019 World Wide Web Conference}{May 13--17, 2019}{San Francisco, CA, USA}
\acmBooktitle{Proceedings of the 2019 World Wide Web Conference (WWW '19), May 13--17, 2019, San Francisco, CA, USA}
\acmPrice{}
\acmDOI{10.1145/3308558.3313419}
\acmISBN{978-1-4503-6674-8/19/05}

\newcommand{\MyParagraph}[1]{\smallskip\indent\emph{#1.} }

\usepackage{hyperref}
\usepackage{booktabs} 
\usepackage{subfig}
\usepackage[skip=0pt]{caption}
\usepackage{graphicx}
\usepackage{paralist}
\usepackage{enumitem}
\usepackage{acronym}
\usepackage{amsmath}
\acrodef{LTR}{learning to rank}
\acrodef{ViTOR}{\textit{Visual learning TO Rank}}
\acrodef{SERP}{search engine results page}

\newcommand{\datasetname}{\ac{ViTOR}}
\newcommand{\modelname}{\ac{ViTOR}}
\newcommand{\modelnamef}{\acf{ViTOR}}
\newcommand{\OK}{@{\mbox{}\hspace*{.25cm}}}

\title{ViTOR: Learning to Rank Webpages Based on Visual Features}

\author{Bram van den Akker}
\orcid{}
\affiliation{%
  \institution{University of Amsterdam}
  \city{Amsterdam} 
  \country{The Netherlands}
}
\email{contact@bramvandenakker.nl}

\author{Ilya Markov}
\orcid{}
\affiliation{%
  \institution{University of Amsterdam}
  \city{Amsterdam} 
  \country{The Netherlands}  
}
\email{i.markov@uva.nl}

\author{Maarten de Rijke}
\orcid{0000-0002-1086-0202}
\affiliation{%
   \institution{University of Amsterdam}
   \city{Amsterdam} 
   \country{The Netherlands}
}
\email{derijke@uva.nl}

\renewcommand{\shortauthors}{B. van den Akker et al.}

\begin{document}

%
%
\begin{abstract}
The visual appearance of a webpage carries valuable information about the page's quality and can be used to improve the performance of \ac{LTR}.
We introduce the \modelname~model that integrates state-of-the-art visual features extraction methods:
\begin{inparaenum}[(i)]
\item transfer learning from a pre-trained image classification model, and
\item synthetic saliency heat maps generated from webpage snapshots.
\end{inparaenum}
Since there is currently no public dataset for the task of \ac{LTR} with visual features, we also introduce and release the \datasetname~dataset, containing visually rich and diverse webpages.
The \datasetname~dataset consists of visual snapshots, non-visual features and relevance judgments for ClueWeb12 webpages and TREC Web Track queries.
We experiment with the proposed \modelname~model on the \datasetname~dataset
and show that it significantly improves the performance of \ac{LTR} with visual features.
\end{abstract}


\maketitle


\section{Introduction}
The design and appearance of a webpage are determining factors for a user to examine the page or to divert to another page~\cite{nielsen1999designing,nielsen2006f,pernice2017f,wang2014eye}.
However, relatively little is known about the potential of visual appearance to help determine the perceived relevance of a webpage.
Recently, visual features, extracted from snapshots of webpages and \acp{SERP}, have been introduced into \acf{LTR}
and have been shown to significantly improve the \ac{LTR} performance~\cite{fan2017learning,zhang2018relevance}.

In this paper we continue studying \ac{LTR} with visual features and propose the \modelnamef~model
that integrates state-of-the-art visual features extraction methods.
We present two implementations of the \modelname~model.
First, we extract visual features from webpage snapshots using transfer learning and, in particular, by adopting the VGG-16~\cite{simonyan2014very} and ResNet-152~\cite{he2016deep} models pre-trained on ImageNet.
Second, we introduce a novel set of visual features extracted from synthetic saliency heatmaps, which explicitly model how users view webpages~\cite{shan2017two}.

Currently, there is no dataset available to support research on \ac{LTR} with visual features.
\citet{fan2017learning} experimented with webpages from the GOV2 collection.\footnote{\url{http://ir.dcs.gla.ac.uk/test_collections/gov2-summary.htm}}
However, GOV2 solely contains webpages within the .gov domain, i.e., pages with a relatively narrow scope,
and, more importantly, these webpages do not contain their original images and styles.
Hence, the GOV2 collection cannot fully support research on \ac{LTR} with visual features, as it does not reflect visually rich and diverse webpages found on the internet today.

To overcome this issue, we release the \datasetname~dataset that contains webpages from the ClueWeb12 collection\footnote{\url{https://lemurproject.org/clueweb12/}}
and queries from the TREC Web Tracks 2013 \& 2014~\cite{collins2013trec,collins2015trec}.
For each webpage of ClueWeb12 that also appears in the web tracks, the \datasetname~dataset contains a snapshot
and a set of content features, such as BM25 and PageRank.
The introduced dataset is made publicly available.\footnote{\url{https://github.com/Braamling/learning-to-rank-webpages-based-on-visual-features/blob/master/dataset.md}}

To assess the performance of the proposed \modelname~model, we run experiments on the introduced \datasetname~dataset.
Our experiments confirm that visual features significantly improve the \ac{LTR} performance, which is inline with previous findings~\cite{fan2017learning}.
We also show that both implementations of the \modelname~model, namely transfer learning (with VGG-16 and ResNet-152) and synthetic saliency heatmaps,
significantly outperform other \ac{LTR} methods with visual features.

In summary, the main contributions of this work are:
\begin{inparaenum}[(i)]
\item We introduce transfer learning to extract visual features for \ac{LTR}.
\item We introduce synthetic saliency heatmaps as a novel set of \ac{LTR} features.
\item We introduce and publish \datasetname, an out-of-the-box dataset for \ac{LTR} with visual features.
\end{inparaenum}

\section{Related work}
\label{sec:relatedwork}


%

In this section, we discuss related research on the usage of visual information in \ac{LTR} and the relation between visual appearance and user perception of webpages.

Using eye-tracking, \citet{nielsen2006f} and \citet{pernice2017f} demonstrate that webpage design and content placement influences the ability of users to find information they are looking for. 
Both studies show that by organizing the content in certain shapes (e.g., an F-shape), information can be navigated more efficiently.
\citet{wang2014eye} show that the size of the fixation areas measured using eye-tracking is larger on webpages with more content, which increases the likelihood that the attention of a user is distracted.
Such studies highlight the importance of the visual appearance of a webpage and its effect on how users perceive pages, demonstrating that visual information has to be taken into account when ranking webpages.

\citet{zhang2018relevance} consider using visual features for \ac{LTR} and, specifically, for learning to re-rank.
The authors propose a multimodal architecture for re-ranking snippets on a \ac{SERP} by learning their visual patterns.
This work demonstrates that combining both visual and non-visual features can improve re-ranking performance.

The closest work to ours is the study by \citet{fan2017learning}, which uses visual information to rank webpages instead of re-ranking snippets within an existing ranking.
The authors use snapshots of webpages to extract visual features for LTR
and show that such visual features significantly improve retrieval performance.
They feed snapshots through a neural network that attempts to model the previously mentioned F-shape.
The output of this neural network is then concatenated with more traditional ranking signals, such as BM25 and PageRank.
Finally, the proposed model (called ViP) is trained end-to-end using a pairwise loss.
In our paper, we continue this line of research and propose the \modelname~model for \ac{LTR} with visual features
that makes use of the state-of-the-art visual feature extraction techniques and shows superior performance compared to ViP.
Also, we develop and publish the \datasetname~dataset that contains visually diverse webpages
compared to the GOV2 dataset used in~\cite{fan2017learning}, which lacks visual diversity and, importantly, does not contain images and style information together with webpages.

\section{\protect\modelname{} Model}
In this section, we introduce the \modelname{} model for \ac{LTR} with visual features.
The proposed model consists of three parts.
First, we introduce the model architecture in Section~\ref{sec:multimodal}.
Then, in Section~\ref{sec:visualfeatures} we describe two visual feature extractors used by \modelname{}:
VGG-16~\cite{simonyan2014very} and ResNet-152~\cite{he2016deep}, both pre-trained on ImageNet.
Finally, in Section~\ref{sec:saliency} we propose to enhance \modelname{} by generating synthetic saliency heatmaps for each of the input images.
 The implementation of the proposed \modelname{} model is available online.\footnote{\label{coderef}\url{https://github.com/Braamling/learning-to-rank-webpages-based-on-visual-features}}

\subsection{Architecture} \label{sec:multimodal}
\begin{figure}[t]
\includegraphics[width = 3.4in]{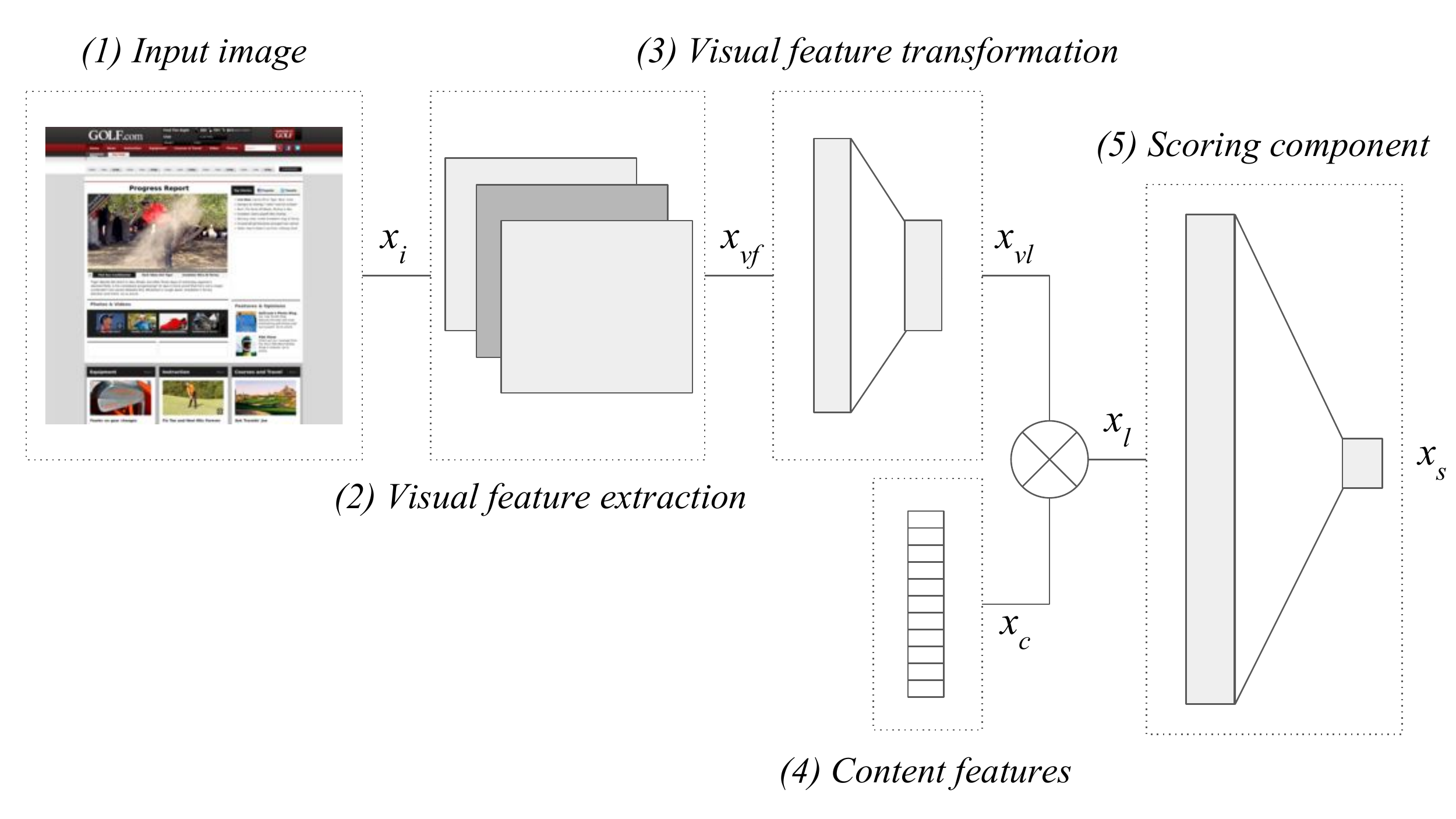}
\caption{\modelname{} architecture.}
\label{fig:multimodelarchitecture}
\end{figure}

The \modelname{} architecture is visualized in Figure~\ref{fig:multimodelarchitecture}. 
The process starts by taking an image $x_i$ (1) as an input to the visual feature extraction layer (2) in order to create a generic visual feature vector $x_{vf}$. 
These features are considered generic because they can be extracted using convolutional filters trained on a different dataset and task. 
In order to transform generic visual features into \ac{LTR} specific visual features, we use $x_{vf}$ as an input to the visual feature transformation layer~(3).
This visual feature transformation layer outputs a visual feature vector $x_{vl}$ that can be used in combination with other \ac{LTR} features. 

Separating the visual feature extraction and transformation layers allows us to significantly reduce the computational requirements when using a pre-trained model for visual feature extraction. 
In Section~\ref{sec:sectionoptimalization} we further elaborate on how the computational requirements can be reduced.
In Section~\ref{sec:visualfeatures} below, we demonstrate how transfer learning can be applied to use pre-trained visual feature extraction methods in combination with the \modelname{} architecture.

The \modelname{} architecture also makes use of content features $x_{c}$~(4), e.g., BM25, PageRank, etc.
The final feature vector $x_{l}$ is constructed by concatenating the visual features $x_{vl}$ with the content features~$x_{c}$.
This final feature vector is then used as an input to the scoring component (5),
which transforms the features into a single score $x_s$ for each query-document pair.
The resulting model is trained end-to-end using a pairwise hinge loss with $L_2$ regularization similarly to~\cite{fan2017learning}.
The scoring component uses a single fully connected layer with a hidden size of $10$ and dropout set to $10\%$,
which showed good performance in preliminary experiments.


\subsection{Visual feature extractors} \label{sec:visualfeatures}
In order to use webpage snapshots in \ac{LTR}, the snapshots have to be converted to a vector representation that can then be used in combination with existing content features. 
Since preparing training data for \ac{LTR} is costly and its amount is usually low, it is beneficial to use a visual feature extraction method that is already pre-trained.
In this work we use the VGG-16~\cite{simonyan2014very} and ResNet-152~\cite{he2016deep} models, two well-explored image classification models which both have an implementation with ImageNet pre-trained parameters.
Below, we describe how these two models are implemented within the \modelname{} architecture.

VGG-16~\cite{simonyan2014very} is commonly used for training transfer-learned models,
because it provides a reasonable trade-off between effectiveness and simplicity~\cite{shan2017two}.
Its architecture consists of a set of convolutional layers and fully connected layers. 
The convolutional layers extract features from an input image, which are then used by the fully connected layers to classify the image. 
The convolutional layers of VGG-16 are generic with respect to an input and task~\citep{donahue2014decaf}
and, thus, can be used as a visual feature extractor within the \modelname{} architecture to create generic visual features $x_{vf}$.
Hence, we use the convolutional layers as is, by freezing all the parameters during training.
Because we do not alter any of the convolutional layers, the size of $x_{vf}$ is determined by the output of the convolutional layers in the original VGG-16 model, being $1\times25088$.

The fully connected layers of VGG-16, instead, can be altered and retrained in order to be used with new inputs and tasks.
Due to this, we utilize them as a visual feature transformation layer within the \modelname{} architecture to produce \ac{LTR} specific features $x_{vl}$.
In particular, we replace the last fully connected layer of VGG-16 by a newly initialized fully connected layer.
Then we optimize the parameters of all fully connected layers of VGG-16 during training.
The size of $x_{vl}$ is set to $30$, as this size showed good performance in preliminary experiments.

The ResNet-152~\cite{he2016deep} architecture was shown to outperform VGG-16 in ImageNet classification.
The residual connections between convolutional layers of ResNet-152 allow for deeper networks to be trained without suffering from vanishing gradients.
Similarly to VGG-16, ResNet-152 has convolutional layers that extract features from an input image, which are in turn used by a fully connected layer to classify each image.
We use these convolutional layers as the visual feature extraction layer, which transforms $x_{i}$ to $x_{vf}$. All parameters of these convolutional layers are frozen during training. As with VGG-16, the size of $x_{vf}$ is determined by the output size of the original convolutional layers in the ResNet-152 model, $1\times2048$.

Additionally, the original ResNet-152 architecture only has a single fully connected layer, which empirically showed to not be enough to successfully train the \modelname~model.
Instead, we transform $x_{vf}$ to $x_{vl}$ by training a fully connected network from scratch.
The transformation layer is constructed using three layers with each having $4096$ hidden units and a final layer resulting in $x_{vl}$ with a size of $30$, which was empirically found to provide good performance in preliminary experiments.

\subsection{Saliency heatmaps} \label{sec:saliency}
\begin{figure}[t]
\begin{tabular}{ccc}
\subfloat{\includegraphics[width = 1in]{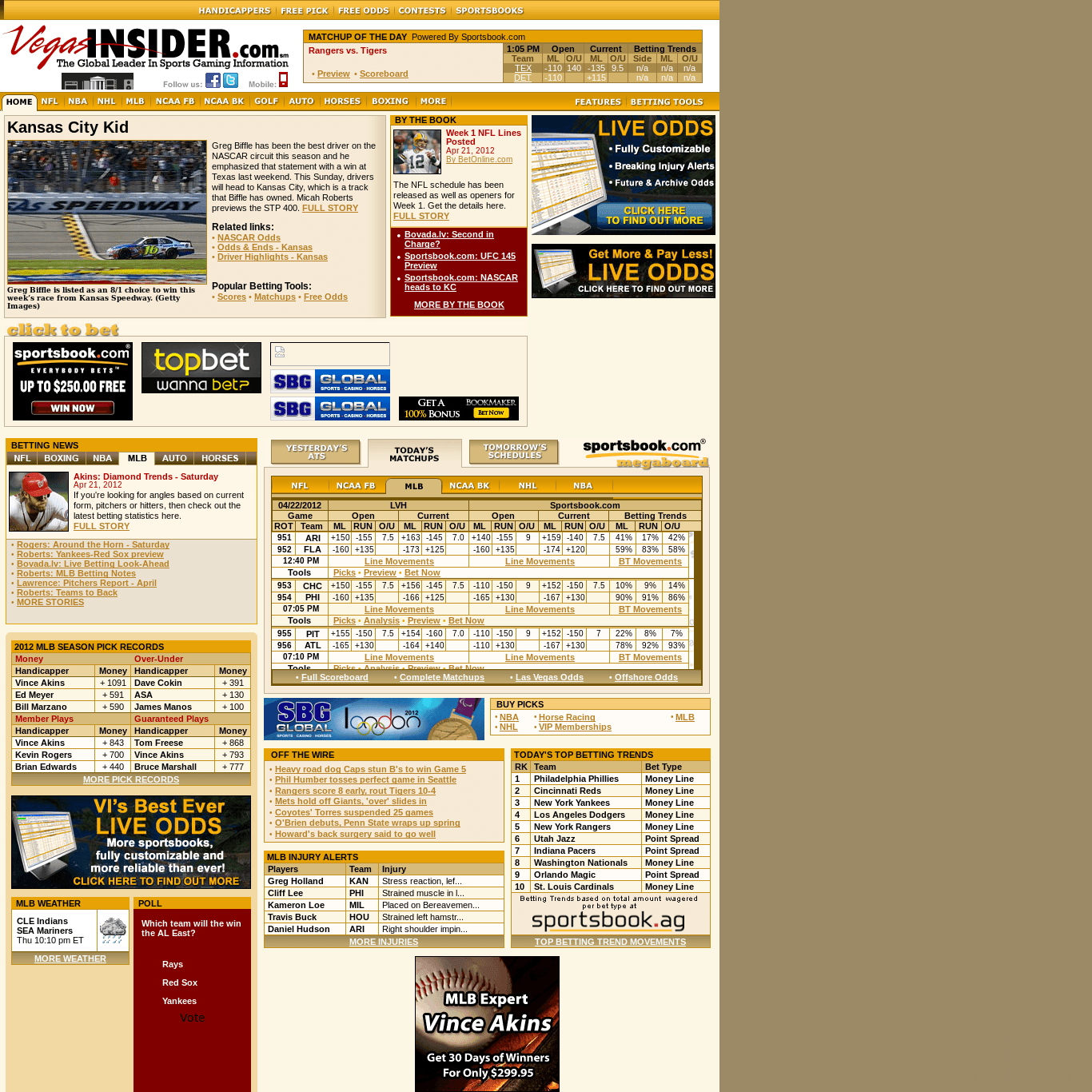}} &
\subfloat{\includegraphics[width = 1in]{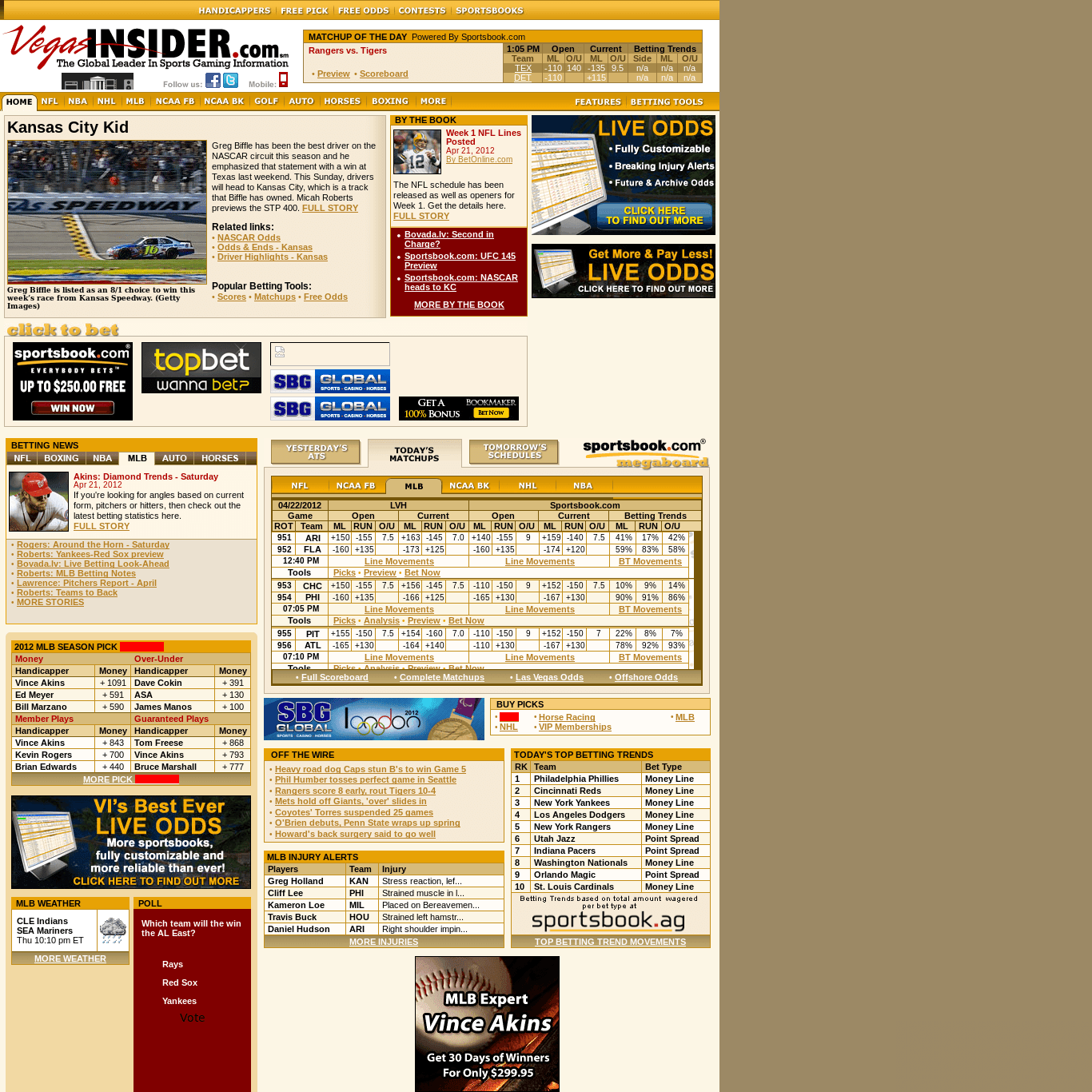}} &
\subfloat{\includegraphics[width = 1in]{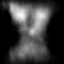}} \\
\subfloat{\includegraphics[width = 1in]{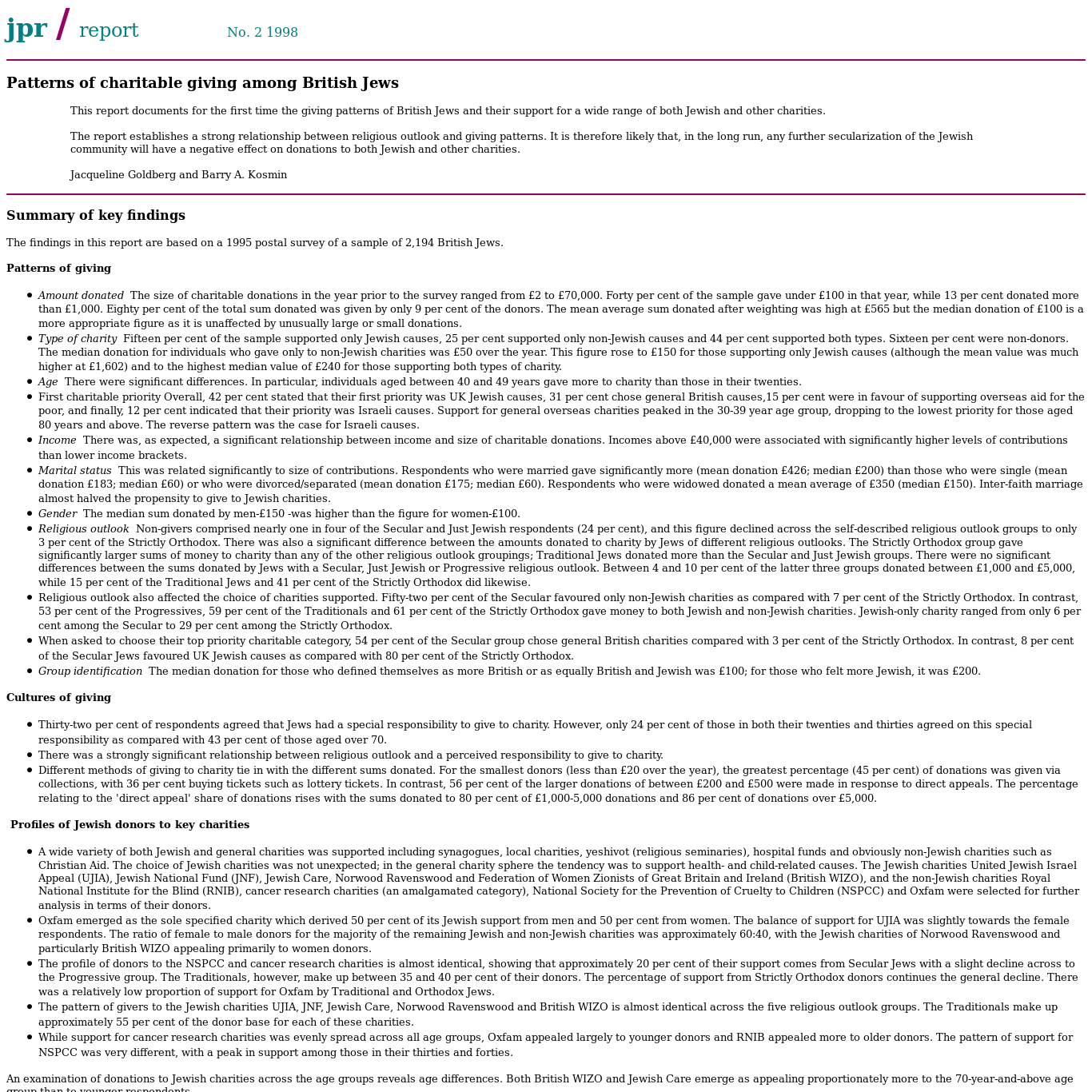}} &
\subfloat{\includegraphics[width = 1in]{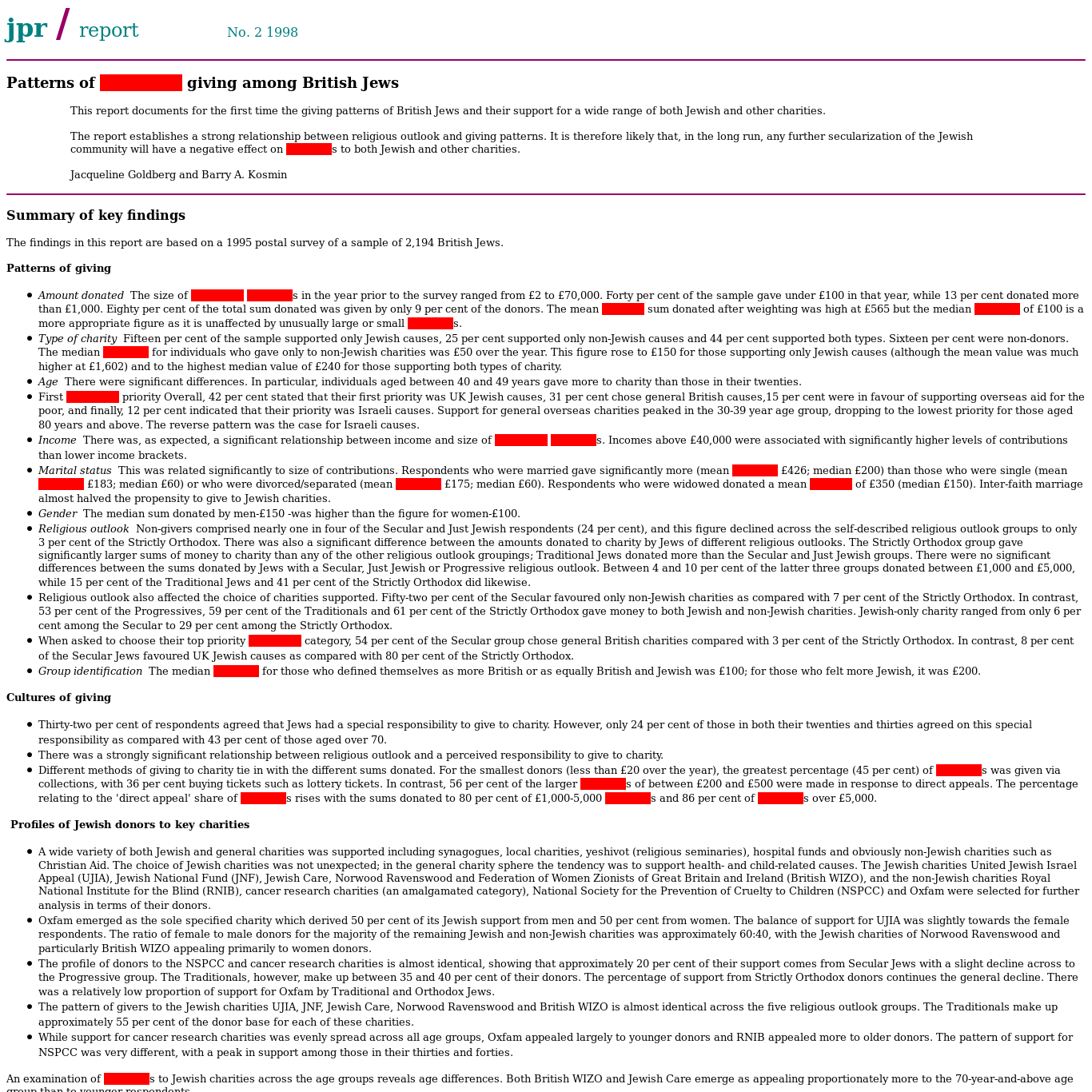}} &
\subfloat{\includegraphics[width = 1in]{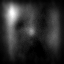}} \\
\subfloat{\includegraphics[width = 1in]{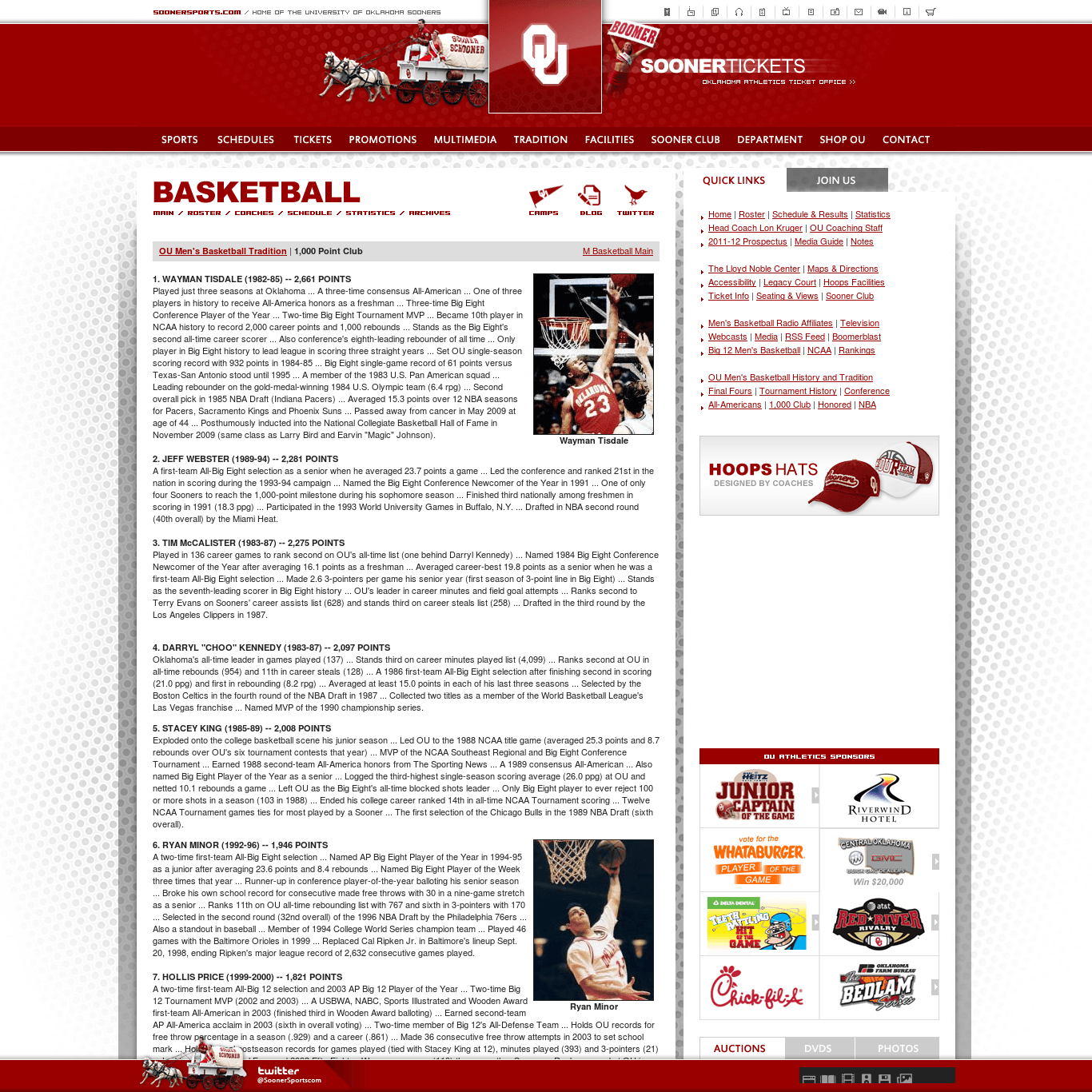}} &
\subfloat{\includegraphics[width = 1in]{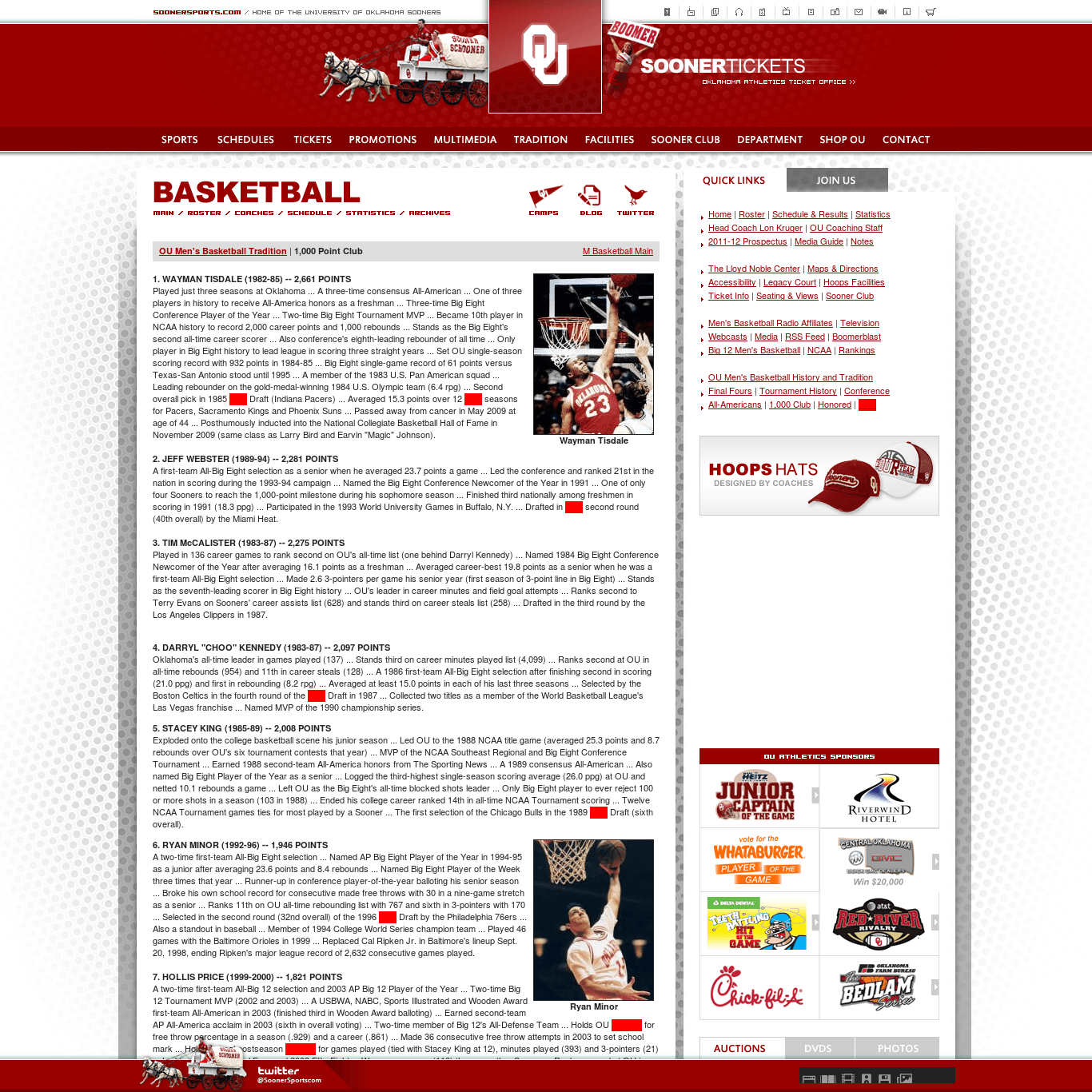}} &
\subfloat{\includegraphics[width = 1in]{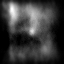}} \\
\end{tabular}
\caption{Examples of a vanilla snapshot, a red highlighted snapshot, and a saliency heatmap from left to right, respectively.}
\label{fig:exampleshots}
\end{figure}

In order to increase the ability to learn the visual quality of a webpage, we propose to explicitly model the user viewing pattern through synthetic saliency heatmaps.
The use of saliency heatmaps could be advantageous compared to the use of raw snapshots for the following reasons.
First, synthetic saliency heatmaps explicitly learn to predict how users perceive webpages by training an end-to-end model on actual eye-tracking data.
We expect this information to better correlate with webpage relevance compared to raw snapshots.
Second, saliency heatmaps reduce the average storage requirements by up to 90\%,
because they are gray-scale images and have large areas of the same color, which can be stored efficiently.
This makes the use of saliency heatmaps attractive for practical applications.
Figure~\ref{fig:exampleshots} shows example snapshots with their corresponding heatmaps (first and third columns respectively).

Following \cite{shan2017two}, we use a two-stage transfer learning model that learns how to predict saliency heatmaps on webpages.
Similarly to the visual feature extraction approaches above, \cite{shan2017two} takes a pre-trained image recognition model and finetunes the output layers on the following two datasets in order respectively:
\begin{inparaenum}[(i)]
\item SALICON~\cite{jiang2015salicon}, a large dataset containing saliency heatmaps created with eye-tracking hardware on natural images, and 
\item the webpage saliency dataset from \cite{shen2014webpage}, a smaller dataset containing saliency heatmaps created with eye-tracking hardware on webpages.
\end{inparaenum}

The trained model is used to convert a raw snapshot into a synthetic saliency heatmap. This heatmap is then used as an input image $x_i$ for the \modelname{} model (see Figure~\ref{fig:multimodelarchitecture}).
%

\section{\protect\datasetname{} Data\-set}
In this section, we introduce the \datasetname~data\-set for \ac{LTR} with visual features. 
Section~\ref{sec:trecclue} contains information about the underlying ClueWeb12 collection and TREC Web Track topics.
Section~\ref{sec:screenshotsec} explains how the snapshots for ClueWeb12 are acquired.
Section~\ref{sec:contentfeature} discusses content features, such as BM25 and TF-IDF, included in the \datasetname{} dataset.
Finally, Section~\ref{sec:finalcollection} gives an overview of the structure in which the \datasetname~dataset is presented and published.

\subsection{ClueWeb12 \& TREC Web Track}\label{sec:trecclue}
For the \datasetname{} dataset we choose to use a combination of the ClueWeb12 document collection and the topics from the TREC Web Tracks 2013 \& 2014~\cite{collins2013trec,collins2015trec},
because this is currently the most recent combination of a large-scale webpage collection together with judged queries (with graded relevance). 

ClueWeb12 is a highly diverse collection of webpages scraped in the first half of 2012.
The total collection contains over 700 million webpages.
%
We only use ClueWeb12 webpages that have judgements for any of the 100 queries in the TREC Web Tracks 2013 \& 2014.
In total, there are 28,906 judged webpages.
Table~\ref{tab:countsources} shows the breakdown of the total number of webpages and different relevance labels in the combined set of topics from 2013 and 2014.

\begin{table}[h]
  \captionof{table}{Number of webpages per source and the corresponding breakdown of TREC Web Track relevance grades.} 
  \label{tab:countsources}
  \begin{tabular}{ l  @{}r  r  r  r }
  \toprule
    Count/Label & TREC Web & Wayback & ClueWeb12 & \mbox{}\hspace*{-.15cm}No image\\
    \midrule
    Total & 28,906 & 23,249 & 5,392 & 265 \\
    Nav grade (4) & 40 & 36 & 4 & 0\\
    Key grade (3) & 409 & 347 & 62 & 0\\
    Hrel grade (2) & 2,534 & 2,222 & 295 & 17 \\
    Rel grade (1) & 6,832 & 5,679 & 1,123 & 30\\
    Non grade (0) & 18,301 & 14,395 & 3,701 & 205 \\
    Junk grade (-2) & 790 & 570 & 207 & 13\\
    \bottomrule
  \end{tabular} 
\end{table}

\subsection{Snapshots} \label{sec:screenshotsec}
Although each entry in the ClueWeb12 collection contains the webpage's HTML source, many pages lack styling and images files in order to render the full page.
To overcome this issue, we use the Wayback Machine,\footnote{\url{http://archive.org/web/}} which offers various archived versions of webpages with styling and images since 2005.
For each page in ClueWeb12, that is also judged in the TREC Web Tracks 2013 \& 2014,
we scrape an entry on the Wayback Machine that is closest to the original page scrape date as recorded in ClueWeb12.
A snapshot is then taken using a headless instance of the Firefox browser.
To reproduce \cite{fan2017learning}, we also create a separate query-dependent dataset with the same snapshots where all query words are highlighted in red (HEX value: \#ff0000).
Examples of snapshots and snapshots with highlights are shown in Figure~\ref{fig:exampleshots} (first two columns).

\label{sec:datasetsum}
Since the Wayback Machine does not contain an archived or working version of each webpage in the ClueWeb12 collection, a filtering process is introduced to maximize the quality of each snapshot. Using the following criteria, a snapshot is selected for each available webpage:
\begin{inparaenum}[(1)]
\item Each webpage is requested from the Wayback Machine. 
\item A webpage that is not on the Wayback Machine, times out, throws a JavaScript error, or results in a PNG snapshot smaller than 100KB is marked as broken. Such webpages are rendered again using the online rendering service provided by ClueWeb12.\footnote{\url{http://boston.lti.cs.cmu.edu/Services/}}
\item A manual selection is made between all webpages that are rendered from both sources. The Wayback version is used if it contains more styling elements and if the content is the same as in the rendering service. Otherwise, the rendering service version is used. 
\end{inparaenum}

As a result, most of the 28,906 judged webpages have a corresponding snapshot from either the Wayback Machine or ClueWeb12 rendering service.
Only 265 documents did not pass the filtering and were discarded.
Table~\ref{tab:countsources}, row one, summarizes the results of the \datasetname{} dataset~acquisition process. The table also shows the distribution of judgment scores among the snapshots that were taken from the Wayback Machine and ClueWeb12 rendering service.

\subsection{Non-visual features} 
\label{sec:contentfeature}
In LTR, documents are ranked based on various types of features, such as content features (e.g., BM25), quality indicators (e.g., PageRank) and behavioral features (e.g., CTR).
In order to use the \datasetname{} dataset to measure the effect of visual features, we also add a set of content features and quality indicators,
listed in Table~\ref{tab:setdescription}.
These 11 features are chosen so that they resemble most informative features of the most recent LETOR 4.0 dataset~\cite{Qin2013:Introducing} and are easy to compute
(see Section~\ref{sec:contentfeatures} for a detailed comparison of different feature sets).
Also note that behavioral features, such as CTR, are not available for the TREC Web Tracks.

The content features, such as TF-IDF and BM25, are computed by doing a full pass over the complete ClueWeb12 collection.
The PageRank scores are taken from the ClueWeb12 Related Data section.\footnote{\url{https://lemurproject.org/clueweb12/related-data.php}}
The following modifications based on the feature transformations described in LETOR are made to stabilize training:
\begin{inparaenum}[(1)]
\item Free parameters $k_1$, $k_3$ and $b$ for BM25 were set to $2.5$, $0$ and $0.8$, respectively. 
\item Since the PageRank scores are usually an order of magnitude smaller than all other scores, we multiply them by $10^5$.
\item The log transformation is applied to each feature.
\item The log-transformed features are normalized per query.  
\end{inparaenum}

\begin{table}[t]
\centering
\captionof{table}{Non-visual features provided with the \datasetname{} dataset.}  \label{tab:setdescription} 
\begin{tabular}{rlrlrl}
\toprule
Id & Description & Id & Description & Id & Description    \\ 
\midrule
1  & Pagerank  & 5  & Content TF-IDF  & 9  & Title IDF   \\
2  & Content length & 6  & Content BM25   & 10 & Title TF-IDF   \\
3  & Content TF  & 7  & Title length & 11 & Title BM25  \\
4  & Content IDF & 8  & Title TF  & & \\
\bottomrule
\end{tabular}
\end{table}

\subsection{Final collection}\label{sec:finalcollection}
In summary, the \datasetname{} dataset contains:
\begin{inparaenum}[(i)]
\item a directory with webpage snapshots (Section~\ref{sec:screenshotsec}), and
\item a set of files with content features divided into folds (Section~\ref{sec:contentfeature}).
\end{inparaenum}
Each snapshot is stored as a PNG file that can be identified by its corresponding ClueWeb12 document id. 
The non-visual features are stored in LETOR formatted files containing the raw, logged and query normalized values.
The query normalized values are randomly split per query into five equal partitions.
These partitions are then used to create five folds, where each fold contains three partitions for training and the remaining two partitions for validation and testing.



\section{Experimental Setup}\label{sec:setup}
In this section, we discuss the configurations of the \modelname~architecture, baselines and metrics used during our experiments.
%

\MyParagraph{\modelname~configurations.}
We experiment with four configurations of the \modelname~architecture.
\modelname~baseline refers to the \modelname~model with only content features.
This configuration is trained by feeding content features into the scoring component directly, without adding any visual features.
\modelname~snapshots and \modelname~highlights use visual features extracted from vanilla snapshots of webpages and from snapshots of webpages with highlighted query terms, respectively.
Finally, \modelname~saliency uses visual features extracted from synthetic saliency heatmaps. The snapshots, highlights and saliency heatmaps for each model respectively are used as the input image (component (1) of Figure~\ref{fig:multimodelarchitecture}.
For the \modelname~configurations with visual features, we experiment with both VGG-16 and ResNet-152 visual feature extraction methods.
The learning rates for VGG-16 and ResNet-152 are set to and $0.0001$ and $0.00005$, respectively. 
Each experimental run is generated using the Adam optimizer~\cite{kingma2014adam} with a batch size of $100$.

\MyParagraph{Baselines}
We compare our proposed \modelname~model to the ViP model by~\citet{fan2017learning}, the only existing \ac{LTR} method that uses visual features.
We train ViP on both vanilla and highlighted snapshots with the resulting configurations being ViP snapshots and ViP highlights, respectively.
Following~\cite{fan2017learning}, we also compare the \modelname~model to a number of content-based ranking methods,
namely BM25 and state-of-the-art \ac{LTR} techniques, such as RankBoost, AdaRank, and LambdaMart.\footnote{The \ac{LTR} methods are taken from \url{https://sourceforge.net/p/lemur/wiki/RankLib}.}

\MyParagraph{Metrics}
To measure the retrieval performance, we use precision and ndcg at $\{1,10\}$ and MAP.
Statistical significance is determined using a two-tailed paired t-test (p-value $\leq 0.05$). 



\section{Results}
\label{sec:results}
In this section, we present experiments that are set out to test the following:
\begin{inparaenum}[(1)]
    \item the \modelname~model improves the \ac{LTR} performance when introducing visual features, 
    \item synthetic saliency heatmaps improve the \ac{LTR} performance when used as an input the \modelname{} model, and
    \item the \modelname~model improves both visual and non-visual state-of-the-art ranking methods.
\end{inparaenum}

\subsection{\modelname~model with VGG-16 and ResNet-152}
In Table~\ref{tab:letorvisresults}, we compare the performance of the \modelname~model when used with and without visual features.
The first row shows the \modelname~baseline, when using only the content features as an input to the scoring component.
The second to fifth rows show the performance of using VGG-16 and ResNet-152 with both vanilla and highlighted snapshots. 
These results clearly show that both VGG-16 and ResNet-152 visual feature extraction methods significantly improve the performance compared to the \modelname~baseline. 

When comparing the results of the \modelname~model with visual features, we observe the following:
\begin{inparaenum}[(i)]
    \item The highest ranking performance is achieved by using VGG-16 on the highlighted snapshots.
    \item For VGG-16, the values of all metrics are consistently better for highlighted snapshots compared to vanilla snapshots, which is in line with the findings of~\cite{fan2017learning} and is to be expected: highlighted snapshots carry more information compared to vanilla snapshots.
\end{inparaenum}
Based on these results, we conclude that the use of visual features in \ac{LTR} significantly improves performance
and that highlighted snapshots should on average be preferred over vanilla snapshots.

\begin{table}[t]
\caption{Results for the \modelname~model using only content features (baseline), vanilla snapshots, highlighted snapshots, and saliency heatmaps.
All results significantly improve over the \modelname~baseline.
Best results are shown in bold.}
\label{tab:letorvisresults}
\centering
\begin{tabular}{l\OK l\OK l\OK l\OK l\OK l}
\toprule
                      & p@1    & p@10  & ndcg@1  & ndcg@10 & MAP   \\ 
\midrule
\modelname~baseline & 0.338  & 0.370 & 0.189   & 0.233   & 0.415 \\ 
\midrule
VGG snapshots      & 0.514 & 0.484 & 0.292 & 0.324 & 0.442 \\ 
ResNet snapshots   & 0.550 & 0.452 & 0.310 & 0.301 & 0.437 \\ 
VGG highlights     & \textbf{0.560} & \textbf{0.520} & 0.323 & \textbf{0.346} & \textbf{0.456} \\ 
ResNet highlights  & 0.530 & 0.463 & 0.305 & 0.312 & 0.440 \\
\midrule
VGG saliency       & 0.554 & 0.453 & 0.310   & 0.302   & 0.422 \\ 
ResNet saliency    & \textbf{0.560} & 0.476 & \textbf{0.333} & 0.321 & 0.442 \\
\bottomrule
\end{tabular}
\vspace*{-.5\baselineskip}
\end{table}

\subsection{\modelname{} model with saliency heat maps}
The last two rows of Table~\ref{tab:letorvisresults} show the performance of the \modelname{} mo\-del when using synthetic saliency heat maps as an input.
The visual features are learned using both VGG-16 and ResNet-152.
In this case, ResNet-152 consistently outperforms VGG-16.
Although the highlighted snapshots with VGG-16 still outperform ResNet-152 with saliency heat maps on p@10, ndcg@10 and MAP, the saliency heat maps with ResNet-152 match and outperform VGG-16 with highlighted snapshots when looking at p@1 and ndcg@1. 
Hence, saliency heat maps should be preferred in applications where early precision is important,
while highlighted snapshots should be used when a high overall performance is needed. 

\subsection{Baseline comparison}
Table~\ref{tab:baseresults} compares the performance of the \modelname~model to BM25, non-visual \ac{LTR} methods and the ViP model by~\citet{fan2017learning}. 
Specifically, the table shows the performance of VGG-16 with highlighted snapshots and of ResNet-152 with synthetic saliency heatmaps, as these are the best-performing variants of the \modelname~model according to Table~\ref{tab:letorvisresults}.
Both methods have a significant performance increase compared to BM25, almost doubling the metrics values in many cases.

When comparing to non-visual \ac{LTR} methods, both \modelname~implementations show consistently better performance.
However, not all metrics are improved significantly.
We attribute this to the fact that, similarly to~\cite{fan2017learning}, the \ac{LTR} component of the \modelname{} model is based on pairwise hinge loss, which is a relatively simple loss function.

Finally, we compare the \modelname~implementations to ViP, the only existing \ac{LTR} method with visual features.
Here, we clearly see that both our implementations significantly outperform ViP on all metrics.
Also note, that ViP loses to two out of three non-visual \ac{LTR} baselines, namely RankBoost and LambdaMart.
We believe this is due to the reason discussed above: ViP uses pairwise hinge loss as the \ac{LTR} component~\cite{fan2017learning}, which may be suboptimal.

The above results show that the proposed \modelname{} model outperforms baselines, whether they are supervised or unsupervised, use visual features or not.
However, to achieve consistent significant improvements compared to the state-of-the-art \ac{LTR} methods, different loss functions within the \modelname{} model have to be investigated.

\begin{table}[t]
\caption{Results for the VGG-16 with highlighted snapshots, ResNet-152 with saliency heatmaps, and baselines.
$\dagger$ indicates a significant decrease in performance compared to VGG highlights and $\ddagger$ indicates a significant decrease in performance compared to both \modelname{} implementations.
Best results are shown in bold.}

\label{tab:baseresults}
\begin{tabular}{l\OK l\OK l\OK l\OK l\OK l}
\toprule
                      & p@1    & p@10  & ndcg@1  & ndcg@10 & MAP   \\
\midrule
BM25                  & 0.300$^\ddagger$  & 0.316$^\ddagger$ & 0.153$^\ddagger$   & 0.188$^\ddagger$   & 0.350$^\ddagger$ \\ 
\midrule
RankBoost             & 0.450  & 0.444 & 0.258   & 0.288$^\dagger$    & 0.427 \\
AdaRank               & 0.290$^\ddagger$   & 0.357$^\ddagger$  & 0.149$^\ddagger$    & 0.227$^\ddagger$    & 0.398 \\
LambdaMart            & 0.470  & 0.420$^\dagger$ & 0.256   & 0.275$^\dagger$    & 0.418 \\ 
\midrule
ViP snapshots         & 0.392$^\ddagger$ & 0.398$^\ddagger$ & 0.217$^\ddagger$   & 0.254$^\ddagger$   & 0.421$^\ddagger$ \\ 
ViP highlights        & 0.418$^\ddagger$  & 0.416$^\ddagger$ & 0.239$^\ddagger$   & 0.269$^\ddagger$   & 0.422$^\ddagger$ \\
\midrule
VGG highlights        & \textbf{0.560}  & \textbf{0.520} & 0.323   & \textbf{0.346}   & \textbf{0.456} \\ 
ResNet saliency       & \textbf{0.560} & 0.476 & \textbf{0.333} & 0.321 & 0.442 \\
\bottomrule
\end{tabular}
\end{table}


\section{Discussion}
\label{sec:discussion}
In this section, we address two practical aspects of the proposed \modelname~model and dataset:
\begin{inparaenum}[(i)]
    \item the number of parameters of the \modelname{} model and corresponding optimizations, and
    \item the performance of content features included in the \datasetname~dataset.
\end{inparaenum}

\subsection{Training and inference optimization} \label{sec:sectionoptimalization}
Both training and inference using deep convolutional networks are generally computationally expensive.
By separating the feature extraction and transformation layers in the proposed \modelname~architecture (see Figure~\ref{fig:multimodelarchitecture}) we allow powerful computational optimizations for both training and inference. 

\MyParagraph{Training optimization} 
Although we freeze the parameters in the visual feature extraction layer (component (2) of Figure~\ref{fig:multimodelarchitecture}) when using a pre-trained model, we still need to compute the output from all the frozen parameters during the forward propagation.
We can avoid the computational cost associated with the forward propagation on the frozen layers by storing the output of the visual feature extraction layer to disk prior to training.
By storing these vectors to disk, we leave only the parameters of the fully connected layers (component (3) of Figure~\ref{fig:multimodelarchitecture}) to be calculated and stored in memory during the forward propagation. 

By applying the above procedure, we can reduce the number of parameters of the \modelname~model.
When using VGG-16, the visual feature extraction layer consists of $14,714,688$ parameters, which is $12.3\%$ of the parameters in the \modelname{} model.
When using ResNet-152, the visual feature extraction layer consists of $58,144,239$ parameters, which is $49.5\%$ of the parameters in the \modelname{} model.

By using the stored output of the visual feature extraction layer $x_{vf}$ instead of an actual image $x_{i}$ as the input of the \modelname~model,
we reduce the size of each input by $84.7\%$ and $98.6\%$ for VGG-16 and ResNet-152 respectively.
This reduction in input size further reduces the memory required for training the model.

\MyParagraph{Real-time inference optimization}
When using \ac{LTR} in a large-scale production environment, the impact of newly introduced \ac{LTR} features (visual features, in our case) on real-time computational requirements is of major concern.

Since both the vanilla snapshots and synthetic saliency heat\-maps are query independent, the output of the visual feature transformation layer (component (3) of Figure~\ref{fig:multimodelarchitecture}) does not change for different document/query pairs. This enables offline inference, leaving only the scoring component to be inferred in real-time.
Since a visual feature vector is has $1\times30$, using the offline inferred feature vector would result in a \ac{LTR} model with $30 + 11 = 41$ features. This increase in parameters is negligible in terms of computational costs.

\subsection{Benchmarking content features} \label{sec:contentfeatures}
In \ac{LTR} research, both the amount and type of considered features vary widely per dataset and study.
The most recent LETOR 4.0 dataset~\cite{Qin2013:Introducing} contains 46 features extracted for webpages from the GOV2 dataset and queries from the million query tracks 2007 and 2008 (MQ2007 and MQ2008).
In the \datasetname~dataset, we use 11 features that are a subset of the 46 features of LETOR (see Table~\ref{tab:setdescription}).
These 11 features are chosen to be both informative and easy to compute.
Here, we compare our subset of 11 features to the full set of 46 features.
The experiments are run on the GOV2 dataset and MQ2007 queries.
The \ac{LTR} methods considered are the same as in Section~\ref{sec:results}, namely RankBoost, AdaRank, and LambdaMart.

The results of running the considered \ac{LTR} methods using both 46 and 11 features are shown in Table~\ref{tab:11vs46}.
From these results, we see that the number of features has a significant effect on the performance of AdaRank.
However, for the best-performing RankBoost and LambdaRank methods the drop in performance is minor when using 11 features instead of 46 features.
This indicates that the chosen 11 features included in the \datasetname~dataset form a reasonable trade-off between effectiveness and computation cost.

\begin{table}[t]
\caption{Comparison of 46 LETOR features and 11 LETOR features that are also used in \datasetname.}
\label{tab:11vs46}
\centering
\begin{tabular}{l@{~~}ccccc}
\toprule
           & p@1  & p@10   & ndcg@1 & ndcg@10 & MAP \\ 
\midrule
RankBoost - 46 & 0.453 & 0.371 & 0.391 & 0.430  & 0.457 \\
RankBoost - 11 & 0.448 & 0.372 & 0.381  & 0.431   & 0.453 \\
\midrule
AdaRank - 46  & 0.420 & 0.360 & 0.367 & 0.424  & 0.449 \\
AdaRank - 11  & 0.385 & 0.287 & 0.364  & 0.394   & 0.386 \\ 
\midrule
LambdaMart - 46 & 0.452 & 0.384 & 0.405 & 0.444  & 0.463 \\
LambdaMart - 11 & 0.448 & 0.380 & 0.397  & 0.443   & 0.455 \\
\bottomrule
\end{tabular}
\end{table}

\section{Conclusion}
In this paper, we considered the problem of \ac{LTR} with visual features.
We proposed the \modelname~model that extracts visual features from webpage snapshots
using transfer learning (specifically, pre-trained VGG-16 and ResNet-152 models) and synthetic saliency heatmaps.
In both cases the extracted visual features significantly improved the \ac{LTR} performance.
We also showed that the proposed \modelname~model significantly outperformed the visual \ac{LTR} baselines.

In addition to the model, we released the \datasetname~dataset, containing visually rich and diverse webpages with corresponding snapshots.
With the \datasetname~dataset it is now possible to comprehensively study \ac{LTR} with visual features.

One direction for future work is the study of state-of-the-art \ac{LTR} methods, such as RankBoost and LambdaMart, within the scoring components of the \modelname~model.
This could improve performance without changing the visual feature extraction and transformation components.
Another promising direction is to combine multiple visual features,
i.e., visual features extracted from vanilla snapshots, snapshots with highlights and saliency heatmaps. Other visual feature extractors, such as the CapsuleNet~\cite{sabour2017dynamic} model, which is able to learn spatial relations in images, might also provide additional performance. 
Other methods of combining visual and textual features might also be worth exploring. 
Future work could also investigate the robustness of this method when using various rendering variations, e.g., different browser, resolutions. 
Finally, performing a more qualitative analysis on which elements cause improvements in \ac{LTR} with visual features would be interesting, e.g., by interpreting the filters in the feature extractor~\cite{olah2018the}. 

\subsection*{Code and data}

Both the \datasetname~dataset\footnote{\url{https://github.com/Braamling/learning-to-rank-webpages-based-on-visual-features/blob/master/dataset.md}} and the code used to run the experiments in this paper\footnote{\url{https://github.com/Braamling/learning-to-rank-webpages-based-on-visual-features}} are available online.

\subsection*{Acknowledgements}
The Spark experiments in this work were carried out on the Dutch national e-infrastructure with the support of SURF Cooperative. Thanks to the Amsterdam Robotics Lab for using their computational resources. Thanks to Jamie Callan and his team for providing access to the online services for ClueWeb12. 

This research was partially supported by
Ahold Delhaize,
the Association of Universities in the Netherlands (VSNU),
and
the Innovation Center for Artificial Intelligence (ICAI).
All content represents the opinion of the authors, which is not necessarily shared or endorsed by their respective employers and/or sponsors.

\bibliographystyle{ACM-Reference-Format}
\bibliography{www2019-visual-ltr} 


\begin{thebibliography}{18}


\ifx \showCODEN    \undefined \def \showCODEN     #1{\unskip}     \fi
\ifx \showDOI      \undefined \def \showDOI       #1{#1}\fi
\ifx \showISBNx    \undefined \def \showISBNx     #1{\unskip}     \fi
\ifx \showISBNxiii \undefined \def \showISBNxiii  #1{\unskip}     \fi
\ifx \showISSN     \undefined \def \showISSN      #1{\unskip}     \fi
\ifx \showLCCN     \undefined \def \showLCCN      #1{\unskip}     \fi
\ifx \shownote     \undefined \def \shownote      #1{#1}          \fi
\ifx \showarticletitle \undefined \def \showarticletitle #1{#1}   \fi
\ifx \showURL      \undefined \def \showURL       {\relax}        \fi
\providecommand\bibfield[2]{#2}
\providecommand\bibinfo[2]{#2}
\providecommand\natexlab[1]{#1}
\providecommand\showeprint[2][]{arXiv:#2}

\bibitem[\protect\citeauthoryear{Collins-Thompson, Bennett, Diaz, Clarke, and
  Voorhees}{Collins-Thompson et~al\mbox{.}}{2013}]%
        {collins2013trec}
\bibfield{author}{\bibinfo{person}{Kevyn Collins-Thompson},
  \bibinfo{person}{Paul Bennett}, \bibinfo{person}{Fernando Diaz},
  \bibinfo{person}{Charlie Clarke}, {and} \bibinfo{person}{Ellen~M Voorhees}.}
  \bibinfo{year}{2013}\natexlab{}.
\newblock \showarticletitle{TREC 2013 Web Track overview}. In
  \bibinfo{booktitle}{{\em TREC}}. \bibinfo{publisher}{NIST}.
\newblock


\bibitem[\protect\citeauthoryear{Collins-Thompson, Macdonald, Bennett, Diaz,
  and Voorhees}{Collins-Thompson et~al\mbox{.}}{2015}]%
        {collins2015trec}
\bibfield{author}{\bibinfo{person}{Kevyn Collins-Thompson},
  \bibinfo{person}{Craig Macdonald}, \bibinfo{person}{Paul Bennett},
  \bibinfo{person}{Fernando Diaz}, {and} \bibinfo{person}{Ellen~M Voorhees}.}
  \bibinfo{year}{2015}\natexlab{}.
\newblock \showarticletitle{TREC 2014 Web Track overview}. In
  \bibinfo{booktitle}{{\em TREC}}. \bibinfo{publisher}{NIST}.
\newblock


\bibitem[\protect\citeauthoryear{Donahue, Jia, Vinyals, Hoffman, Zhang, Tzeng,
  and Darrell}{Donahue et~al\mbox{.}}{2014}]%
        {donahue2014decaf}
\bibfield{author}{\bibinfo{person}{Jeff Donahue}, \bibinfo{person}{Yangqing
  Jia}, \bibinfo{person}{Oriol Vinyals}, \bibinfo{person}{Judy Hoffman},
  \bibinfo{person}{Ning Zhang}, \bibinfo{person}{Eric Tzeng}, {and}
  \bibinfo{person}{Trevor Darrell}.} \bibinfo{year}{2014}\natexlab{}.
\newblock \showarticletitle{Decaf: A deep convolutional activation feature for
  generic visual recognition}. In \bibinfo{booktitle}{{\em ICML}}.
  \bibinfo{pages}{647--655}.
\newblock


\bibitem[\protect\citeauthoryear{Fan, Guo, Lan, Xu, Pang, and Cheng}{Fan
  et~al\mbox{.}}{2017}]%
        {fan2017learning}
\bibfield{author}{\bibinfo{person}{Yixing Fan}, \bibinfo{person}{Jiafeng Guo},
  \bibinfo{person}{Yanyan Lan}, \bibinfo{person}{Jun Xu},
  \bibinfo{person}{Liang Pang}, {and} \bibinfo{person}{Xueqi Cheng}.}
  \bibinfo{year}{2017}\natexlab{}.
\newblock \showarticletitle{Learning visual features from snapshots for web
  search}. In \bibinfo{booktitle}{{\em CIKM}}. ACM, \bibinfo{pages}{247--256}.
\newblock


\bibitem[\protect\citeauthoryear{He, Zhang, Ren, and Sun}{He
  et~al\mbox{.}}{2016}]%
        {he2016deep}
\bibfield{author}{\bibinfo{person}{Kaiming He}, \bibinfo{person}{Xiangyu
  Zhang}, \bibinfo{person}{Shaoqing Ren}, {and} \bibinfo{person}{Jian Sun}.}
  \bibinfo{year}{2016}\natexlab{}.
\newblock \showarticletitle{Deep residual learning for image recognition}. In
  \bibinfo{booktitle}{{\em CVPR}}. \bibinfo{pages}{770--778}.
\newblock


\bibitem[\protect\citeauthoryear{Jiang, Huang, Duan, and Zhao}{Jiang
  et~al\mbox{.}}{2015}]%
        {jiang2015salicon}
\bibfield{author}{\bibinfo{person}{Ming Jiang}, \bibinfo{person}{Shengsheng
  Huang}, \bibinfo{person}{Juanyong Duan}, {and} \bibinfo{person}{Qi Zhao}.}
  \bibinfo{year}{2015}\natexlab{}.
\newblock \showarticletitle{Salicon: Saliency in context}. In
  \bibinfo{booktitle}{{\em CVPR}}. \bibinfo{pages}{1072--1080}.
\newblock


\bibitem[\protect\citeauthoryear{Kingma and Ba}{Kingma and Ba}{2014}]%
        {kingma2014adam}
\bibfield{author}{\bibinfo{person}{Diederik~P Kingma} {and}
  \bibinfo{person}{Jimmy Ba}.} \bibinfo{year}{2014}\natexlab{}.
\newblock \showarticletitle{Adam: A method for stochastic optimization}.
\newblock \bibinfo{journal}{{\em arXiv preprint arXiv:1412.6980\/}}
  (\bibinfo{year}{2014}).
\newblock


\bibitem[\protect\citeauthoryear{Nielsen}{Nielsen}{1999}]%
        {nielsen1999designing}
\bibfield{author}{\bibinfo{person}{Jakob Nielsen}.}
  \bibinfo{year}{1999}\natexlab{}.
\newblock \bibinfo{booktitle}{{\em Designing web usability: The practice of
  simplicity}}.
\newblock \bibinfo{publisher}{New Riders Publishing}.
\newblock


\bibitem[\protect\citeauthoryear{Nielsen}{Nielsen}{2006}]%
        {nielsen2006f}
\bibfield{author}{\bibinfo{person}{Jakob Nielsen}.}
  \bibinfo{year}{2006}\natexlab{}.
\newblock \bibinfo{title}{F-shaped pattern for reading web content}.
\newblock \bibinfo{howpublished}{Jakob Nielsen's Alertbox}.
  (\bibinfo{year}{2006}).
\newblock


\bibitem[\protect\citeauthoryear{Olah, Satyanarayan, Johnson, Carter, Schubert,
  Ye, and Mordvintsev}{Olah et~al\mbox{.}}{2018}]%
        {olah2018the}
\bibfield{author}{\bibinfo{person}{Chris Olah}, \bibinfo{person}{Arvind
  Satyanarayan}, \bibinfo{person}{Ian Johnson}, \bibinfo{person}{Shan Carter},
  \bibinfo{person}{Ludwig Schubert}, \bibinfo{person}{Katherine Ye}, {and}
  \bibinfo{person}{Alexander Mordvintsev}.} \bibinfo{year}{2018}\natexlab{}.
\newblock \showarticletitle{The Building Blocks of Interpretability}.
\newblock \bibinfo{journal}{{\em Distill\/}} (\bibinfo{year}{2018}).
\newblock


\bibitem[\protect\citeauthoryear{Pernice}{Pernice}{2017}]%
        {pernice2017f}
\bibfield{author}{\bibinfo{person}{Kara Pernice}.}
  \bibinfo{year}{2017}\natexlab{}.
\newblock \bibinfo{title}{F-shaped pattern of reading on the web:
  Misunderstood, but still relevant (Even on mobile)}.
\newblock \bibinfo{howpublished}{nngroup.com}.   (\bibinfo{year}{2017}).
\newblock


\bibitem[\protect\citeauthoryear{Qin and Liu}{Qin and Liu}{2013}]%
        {Qin2013:Introducing}
\bibfield{author}{\bibinfo{person}{Tao Qin} {and} \bibinfo{person}{Tie{-}Yan
  Liu}.} \bibinfo{year}{2013}\natexlab{}.
\newblock \showarticletitle{Introducing {LETOR} 4.0 Datasets}.
\newblock \bibinfo{journal}{{\em arXiv preprint arXiv:1306.2597\/}}
  (\bibinfo{year}{2013}).
\newblock


\bibitem[\protect\citeauthoryear{Sabour, Frosst, and Hinton}{Sabour
  et~al\mbox{.}}{2017}]%
        {sabour2017dynamic}
\bibfield{author}{\bibinfo{person}{Sara Sabour}, \bibinfo{person}{Nicholas
  Frosst}, {and} \bibinfo{person}{Geoffrey~E Hinton}.}
  \bibinfo{year}{2017}\natexlab{}.
\newblock \showarticletitle{Dynamic routing between capsules}. In
  \bibinfo{booktitle}{{\em NIPS}}. \bibinfo{pages}{3859--3869}.
\newblock


\bibitem[\protect\citeauthoryear{Shan, Sun, Zhou, and Liu}{Shan
  et~al\mbox{.}}{2017}]%
        {shan2017two}
\bibfield{author}{\bibinfo{person}{Wei Shan}, \bibinfo{person}{Guangling Sun},
  \bibinfo{person}{Xiaofei Zhou}, {and} \bibinfo{person}{Zhi Liu}.}
  \bibinfo{year}{2017}\natexlab{}.
\newblock \showarticletitle{Two-stage transfer learning of end-to-end
  convolutional neural networks for webpage saliency prediction}. In
  \bibinfo{booktitle}{{\em IScIDE}}. Springer, \bibinfo{pages}{316--324}.
\newblock


\bibitem[\protect\citeauthoryear{Shen and Zhao}{Shen and Zhao}{2014}]%
        {shen2014webpage}
\bibfield{author}{\bibinfo{person}{Chengyao Shen} {and} \bibinfo{person}{Qi
  Zhao}.} \bibinfo{year}{2014}\natexlab{}.
\newblock \showarticletitle{Webpage saliency}. In \bibinfo{booktitle}{{\em
  ECCV}}. Springer, \bibinfo{pages}{33--46}.
\newblock


\bibitem[\protect\citeauthoryear{Simonyan and Zisserman}{Simonyan and
  Zisserman}{2014}]%
        {simonyan2014very}
\bibfield{author}{\bibinfo{person}{Karen Simonyan} {and}
  \bibinfo{person}{Andrew Zisserman}.} \bibinfo{year}{2014}\natexlab{}.
\newblock \showarticletitle{Very deep convolutional networks for large-scale
  image recognition}.
\newblock \bibinfo{journal}{{\em arXiv preprint arXiv:1409.1556\/}}
  (\bibinfo{year}{2014}).
\newblock


\bibitem[\protect\citeauthoryear{Wang, Yang, Liu, Cao, and Ma}{Wang
  et~al\mbox{.}}{2014}]%
        {wang2014eye}
\bibfield{author}{\bibinfo{person}{Qiuzhen Wang}, \bibinfo{person}{Sa Yang},
  \bibinfo{person}{Manlu Liu}, \bibinfo{person}{Zike Cao}, {and}
  \bibinfo{person}{Qingguo Ma}.} \bibinfo{year}{2014}\natexlab{}.
\newblock \showarticletitle{An eye-tracking study of website complexity from
  cognitive load perspective}.
\newblock \bibinfo{journal}{{\em Decision Support Systems\/}}
  \bibinfo{volume}{62} (\bibinfo{year}{2014}), \bibinfo{pages}{1--10}.
\newblock


\bibitem[\protect\citeauthoryear{Zhang, Liu, Ma, and Tian}{Zhang
  et~al\mbox{.}}{2018}]%
        {zhang2018relevance}
\bibfield{author}{\bibinfo{person}{Junqi Zhang}, \bibinfo{person}{Yiqun Liu},
  \bibinfo{person}{Shaoping Ma}, {and} \bibinfo{person}{Qi Tian}.}
  \bibinfo{year}{2018}\natexlab{}.
\newblock \showarticletitle{Relevance Estimation with Multiple Information
  Sources on Search Engine Result Pages}. In \bibinfo{booktitle}{{\em CIKM}}.
  ACM, \bibinfo{pages}{627--636}.
\newblock


\end{thebibliography}

\end{document}